\documentclass[aps,pra,twocolumn,showpacs,amsmath,amssymb]{revtex4-1}

\usepackage{graphicx}
\usepackage{color}
\usepackage{epstopdf}
\usepackage{amsmath}
\usepackage{bm}
\usepackage{hyperref}
\usepackage{cleveref}
\usepackage{tcolorbox}
\usepackage{braket}
\usepackage{physics}
\usepackage[export]{adjustbox}
\usepackage[hang]{subfigure}

\begin{document}

\title{
 Accessing elusive two-dimensional phases of dipolar Bose-Einstein condensates by finite temperature
}

	\author{Liang-Jun He$^{1}$}
 	\author{Juan Sanchez-Baena$^{2}$}
	\author{Fabian Maucher$^{3}$}
	\author{Yong-Chang Zhang$^{1}$}
	\email{zhangyc@xjtu.edu.cn}
	\affiliation{$^1$MOE Key Laboratory for Nonequilibrium Synthesis and Modulation of Condensed Matter, Shaanxi Key Laboratory of Quantum Information and Quantum Optoelectronic Devices, School of Physics, Xi’an Jiaotong University, Xi’an 710049, People’s Republic of China \\
    $^2$Departament de F\'isica, Universitat Polit\`ecnica de Catalunya, Campus Nord B4-B5, 08034 Barcelona, Spain \\
	$^3$Faculty of Mechanical Engineering; Department of Precision and Microsystems Engineering, Delft University of Technology, 2628 CD, Delft, The Netherlands}

\begin{abstract}

It has been shown that dipolar Bose-Einstein condensates that are tightly trapped along the polarization direction can feature a rich phase diagram.
In this paper we show that finite temperature can assist in accessing parts of the phase diagram that otherwise appear hard to realise due to excessively large densities and number of atoms being required.
These include honeycomb and stripe phases both in the thermodynamic limit as well as with a finite extent using both variational and numerical calculations.
We account for the effect of thermal fluctuations by means of
Bogoliubov theory employing the local density approximation.
Furthermore, we exhibit real-time evolution simulations leading to such states.
We find that finite temperatures can lead to a significant decrease in the necessary particle number and density that might ultimately pave a route for future experimental realisations.

\end{abstract}

\maketitle

\section{Introduction}
Dipolar Bose-Einstein condensates (dBECs) represent an outstanding platform for exploring the interplay between long-ranged dipole-dipole interaction, contact interaction and quantum fluctuations~\cite{Chomaz_2023}. Quantum fluctuations~\cite{pelster12,Lima:2012hr} can play a crucial role in stabilising the condensate against collapse~\cite{Koch2008,Lahaye:PRL:2008,Petrov:2015kh,wachtler2016}, providing access to parameter domains where exotic physics occur. This mainly includes self-organised pattern-formation akin to classical ferrofluids~\cite{Rosensweig}, and the emergence of ultradilute liquid droplets~\cite{Pfau:nature:2016,Pfau:nature2:2016,Pfau:PRL:2016,ferlaino16}.

Supersolidity is a state of matter that simultaneously features both discrete translational symmetry and a large superfluid fraction~\cite{gross:1957,Andreev:JETP:1969,Chester:PRA:1970,Leggett:PRL:1970}. Such phase-coherent density-modulated states were realised using additional external fields~\cite{Donner:nature:2017,Ketterle:nature:2017}. With the pioneering experiment of Ref.~\cite{Pfau:nature:2016}, dBECs emerged as self-organising alternative.
Since then, a range of exciting experiments explored pattern-formation and supersolidity in dBECs, including their excitation spectra~\cite{Guo:Nature:2019,Tanzi:Nature:2019,Natale:PRL:2019,Poli:arxiv:2024,blakie2024:arxiv}, nucleation of vortices~\cite{casotti202} and the emergence of patterns in elongated cigar-shaped traps with one-dimensional symmetry breaking~\cite{Modugno:PRL:2019,Pfau:PRX:2019,Ferlaino:PRX:2019,Tanzi:Nature:2019,Guo:Nature:2019,Tanzi:Science:2021,BiagioniPRX2022,Ferlaino:PRL:2021}
and in a pancake trapping geometry, where the condensate is tightly confined along the polarization direction leading to a two-dimensionally broken symmetry~\cite{Pfau:nature:2016,norcia21:nature}. This intense experimental activity has been complemented with a range of theoretical works exploring the physics of dBECs in cigar-shaped~\cite{edler2017,Ancilotto:PRA:2019,Blakie_2020,ferlaino20,Matveenko2022,smith22,blakie2023:PRR,alana24,baena22,baena24} 
and pancake traps~\cite{zhang19,Bland2022,youngs2023,zhang21,norcia:prl:2022,hertkorn21,zhang24}.

Theoretical works exploring such pancake geometries at zero temperature revealed a rich phase-diagram in this system~\cite{zhang19,zhang21,hertkorn21,zhang24} with interesting supersolid properties~\cite{zhang19,gallemi22}. It was shown that the different co-existing phases converge to a single point, at which the phase-transition is of second order and around which supersolidity occurs~\cite{zhang19}. This second-order point unfortunately requires large densities that appear experimentally unfeasible, and the new phases, namely stripe and honeycomb states, require even higher densities. Thus, finding parameters that permit the experimental realisation of these phases represents a significant challenge~\cite{zhang21,hertkorn21}.

Yet, there appears an alternative way to promote roton-softening and, subsequently, drive the quantum phase-transition apart from only increasing density and changing trapping parameters. Given that dBECs are strongly susceptible to quantum fluctuations, it might seem plausible that thermal fluctuations have a strong effect as well.
In fact, recent experiments explored the effect of finite temperatures on the dBEC~\cite{Ferlaino:PRL:2021}. Later, theoretical considerations that treated thermal fluctuations by means of Bogoliubov theory employing local density approximation showed that increasing temperature can indeed promote pattern-formation and possibly supersolidity~\cite{baena22,baena24}.

Here, we pursue the idea whether a finite temperature might lower the density required for accessing the second-order point and the high density phases to experimentally more accessible values for a dBEC in a pancake trap. The shift in density due to a finite temperature has already been explored in a cigar-shaped trap~\cite{baena24}. The motivation for exploring the latter again in a pancake-trap is due to the fact that dimensionality and confinement play a crucial role for the density and particle number at the second-order point~\cite{zhang19,zhang21}.
Thus, we address these questions in this paper, which is organised as follows: In Section II we review the model we employ to describe the finite temperature effects. In Section III we present the finite temperature phase-diagram. In Section IV we explore whether a real-time evolution accounting for three-body losses can actually
drive the transition to the high density phases. Finally, in Section V we present the main conclusions of our work.

\section{\label{sec:finite_T}Finite temperature theory}

To identify ground states of the condensate we use both numerical as well as variational methods.
We employ Bogoliubov theory and the local density approximation to account for  thermal fluctuations on the condensate~\cite{oktel19,oktel20,baena22}.
This leads to the temperature dependent extended Gross-Pitaevskii equation (TeGPE)
for the condensate wave function $\psi({\bf r})$
given by
\begin{align}
 \mu \psi({\bf r}) =& \bigg(-\frac{\hbar^2\nabla^2}{2m} + U({\bf r}) + \!\!\int \!{\rm d}{\bf r}^\prime V_{\rm dd}({\bf r}-{\bf r}') \abs{\psi({\bf r}')}^2+ \nonumber\\
 &+ \frac{4\pi\hbar^2 a_{\rm s}}{m} \abs{\psi({\bf r})}^2 + H_{\rm qu}({\bf r}) + H_{\rm th}({\bf r}) \bigg) \psi({\bf r}) \label{TeGPE} \ .
\end{align}
Here, $\mu$ is the chemical potential, $m$ is the atomic mass, $V_{\rm dd}$ denotes the dipole-dipole interaction and $a_{\rm s}$ is the s-wave scattering length. $U$ describes the trapping potential, which in the thermodynamic limit reads $U({\bf r}) = \frac{1}{2} m \omega_z^2 z^2$ and for the fully trapped system is given by $U({\bf r}) = \frac{1}{2} m \left( \omega_x^2 x^2 + \omega_y^2 y^2 + \omega_z^2 z^2 \right)$ with $\omega_z \gg \omega_x \text{, }\omega_y$. The terms $H_{\rm qu}$ and $H_{\rm th}$ account for the effect of quantum and thermal fluctuations, respectively. They are given by~\cite{pelster12,oktel19,baena22}
\begin{align}
 H_{\rm qu}({\bf r}) &= \frac{32}{3 \sqrt{\pi}} g \sqrt{a_{\rm s}^3} Q_5(a_{\rm dd}/a_{\rm s}) \abs{ \psi({\bf r}) }^3 \label{QF} \\
 H_{\rm th}({\bf r}) &= { \int \frac{d{\bf k}}{(2 \pi)^3} \frac{ 1 }{ \left( e^{\beta \varepsilon_{{\bf k}}}-1 \right) } \tilde{V}({\bf k}) \frac{ \tau_{ {\bf k} } }{ \varepsilon_{\bf k}({\bf r}) } } \label{TF} \ ,
\end{align}
where $g=\frac{4\pi \hbar^2 a_{\rm s}}{m}$, and $\varepsilon_{{\bf k}}({\bf r}) = \sqrt{ \tau_{ {\bf k} } \left( \tau_{ {\bf k} } + 2 |\psi({\bf r})|^2 \tilde{V}({\bf k}) \right) }$ is the Bogoliubov excitation spectrum for a given local density $|\psi({\bf r})|^2$ of the dBEC, $\tau_{ {\bf k} } = \frac{\hbar^2 k^2}{2m}$, $\beta = 1/k_{\rm B} T$ and $T$ denotes temperature. $\tilde{V}({\bf k})$ corresponds to the Fourier transform of the sum of the dipole-dipole interaction and the  contact interaction, given by
\begin{equation}
\tilde{V}( {\bf k} ) = \frac{4 \pi \hbar^2 a_{\rm s}}{m} + \frac{4 \pi \hbar^2 a_{\text{dd}}}{m} \left( 3 \frac{k_z^2}{k^2} -1\right) \ ,
\label{fourier_int}
\end{equation}
where dipoles are assumed to be polarized along the $z$ axis.
The parameter $a_{\rm dd} = m C_{\text{dd}}/(12 \pi \hbar^2)$ corresponds to the dipole length, $C_{\text{dd}}$ describes the strength of the dipolar interaction, and the auxiliary function $Q_5(a_{\rm dd}/a_{\rm s})$ is given by~\cite{pelster12}
\begin{align}
 Q_5(a_{\rm dd}/a_{\rm s}) = \int_{0}^1 du \left( 1 - \frac{a_{\rm dd}}{a_{\rm s}} + 3 \left( \frac{a_{\rm dd}}{a_{\rm s}} \right) u^2 \right)^{5/2} \ .
\end{align}
Eq.~(\ref{QF}) describes quantum fluctuations and is responsible for arresting collapse~\cite{wachtler2016} of the condensate that would otherwise occur~\cite{Koch2008,Lahaye:PRL:2008}. Care must be taken in the evaluation of Eq.~(\ref{TF}), since imaginary excitation energies arise for $a_{\rm s}<a_{\rm dd}$ at low momenta. The application of the trapping potential in all three spatial dimensions implies a finite size of our system in a given trapping direction which provides a lower bound to the possible momenta of the excitations entering Eq.~(\ref{TF}). Due to the symmetry of the dipole-dipole interaction, the contribution to the fluctuation energies depends only on $k_z$ and $k_{\rho} = \sqrt{k_x^2 + k_y^2}$. Thus, we only retain excitations that fulfill $k_z>2\pi/l_z$ and $k_\rho > \sqrt{ (2\pi/l_x)^2 + (2\pi/l_y)^2 }$, with $l_{x,y,z}$ representing the size of our system along the $x,y,z$ axes, respectively. For the homogeneous system, we set $k_z >2\pi/l_z$, $k_{\rho}>0$. In our calculations, we assume that the condensate exhibits a Thomas-Fermi profile with a typical width $\sigma_z=\left(\frac{\rho_{\rm 2D}(a_{\rm s}/a_{\rm dd}+2)}{2\omega^2_z}\right)^{1/3}$ in the $z$ direction~\cite{zhang19,zhang21,zhang23} (also see the subsequent discussion on variational approximation), and we approximately set $l_z=2\sigma_z$. For the transverse size in the fully trapped situation, we first obtain a stable solution to Eq.~(\ref{TeGPE}) using imaginary-time evolution without a transverse cutoff (i.e., $k_\rho>0$). Subsequently, we fit the transverse density profile with a Gaussian function characterized by a width $\sigma_\perp$, which allows us to determine $l_x=l_y=2\sigma_{\perp}$ and thus establish the transverse cutoff $k_\rho>\pi/\sigma_\perp$. Using this finite cutoff, we recalculate the ground state of Eq.~(\ref{TeGPE}).
Note, that the cutoff may slightly alter the exact position of the parameter domains~\cite{zhang:arxiv:24}.
For the variational approximation, we consider the energy difference $\Delta E=E(\rho)-E(\rho_0)$ between the unmodulated state $\rho_0$ and a modulated state $\rho$ containing periodic density perturbations as below~\cite{zhang19,zhang21,zhang23},
\begin{equation}
	\rho(\mathbf{r})=\rho_0(z)\left(1+P(\mathbf{r}_\perp)\right)
	\label{eq:variation}
\end{equation}
where the unmodulated state is approximated by a Thomas-Fermi profile $\rho_0(z)=\frac{3\rho_{\rm 2D}}{4\sigma_z}\left(1-\frac{z^2}{\sigma^2_z} \right)$ 
in the confined $z$ direction, $\rho_{\rm 2D}$ represents the average 2D density in the transverse direction, and $P(\mathbf{r}_\perp)$ describes the periodic perturbation in the transverse $xy$ plane. For the modulation exhibiting three-fold rotational symmetry, we define the density as $P(\mathbf{r}_\perp)=A\sum_{j=1}^3\cos(\mathbf{p}_j\cdot \mathbf{r}_\perp)$, where $A$ represents the modulation amplitude and the three wave vectors $\mathbf{p}_j$ form an equilateral triangle in the transverse direction, satisfying $\sum_{j=1}^3\mathbf{p}_j=0$ and $|\mathbf{p}_j|=p$. In this scenario, Eq.~(\ref{eq:variation}) reveals two distinct density distributions depending on the sign of the modulation amplitude: a triangular state for positive $A$ and a honeycomb state for negative $A$, as discussed in subsequent sections. Similarly, for the modulated state with two-fold rotational symmetry, such as the stripe phase, the density modulation can be expressed as $P(\mathbf{r}_\perp)=A\cos(\mathbf{p}\cdot \mathbf{r}_\perp)$,  which involves only one wave vector component. By substituting this ansatz into the energy difference equation, we obtain the energy difference $\Delta E(A,p)$ as a function of $A$ and $p$. By numerically minimizing $\Delta E$ with respect to the two variational parameters, one can determine the ground state with the lowest energy. A non-negative $\Delta E$ for arbitrary $A$ and $p$ indicates an unmodulated superfluid ground state, while a negative $\Delta E$ at finite $A$ and $p$ indicates the transition to a modulated state. 
By comparing the energy shifts of different types of modulated states, we can identify the boundaries between the triangular, stripe and honeycomb states.

To present the results of our work, we choose the characteristic length and energy scales given by $r_0 = 12 \pi a_{\rm dd}$ and $\epsilon_{\rm dd} = \hbar^2/(m r_0^2)$. Therefore, and if not specified otherwise, all subsequent length and energy scales are expressed in terms of these characteristic quantities.

\section{Finite temperature phase diagram}

\begin{figure}[!t]
	\centering
	\includegraphics[width=\columnwidth]{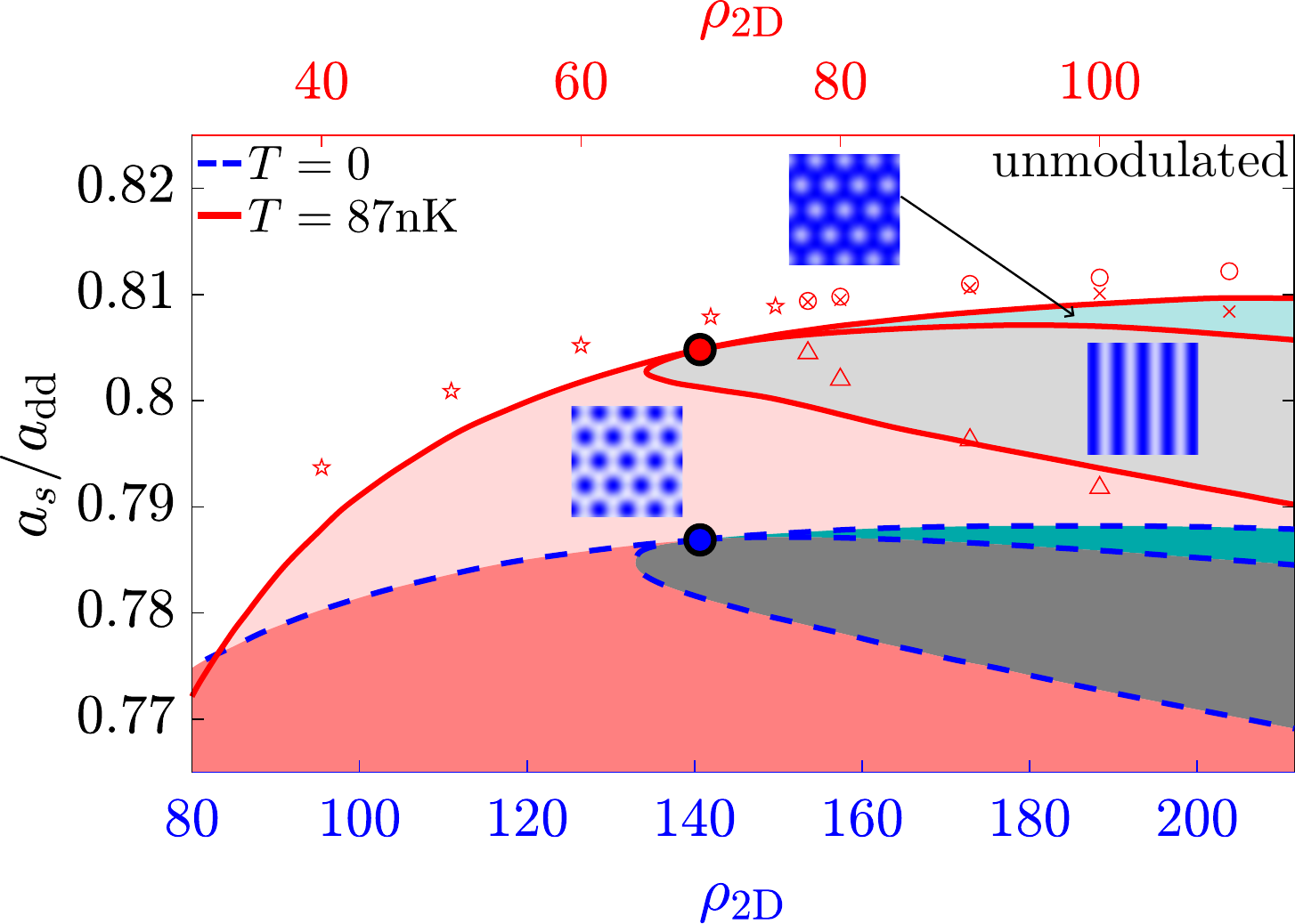}
	\caption{Variational phase diagram for $T=0$ (blue dashed lines) and for $k_{\rm B} T/\epsilon_{\rm dd}=2$ (solid red line) of a pancake dipolar BEC in the thermodynamic limit. The red markers represent the corresponding critical points between different phases obtained via numerically solving the TeGPE. The density profiles of the modulated states (i.e., triangular, stripe, and honeycomb) are displayed in their corresponding domains. The density value is indicated by the color depth, where blue (white) corresponds to a large (low) density. Note, that there is a difference in scale for the 2D condensed density $\rho_{\rm 2D}$ shown at the bottom (i.e., blue axis) for the zero temperature case and at the top (i.e.,red axis) for the finite temperature case. The density is expressed in units of $1/r_0^2$ with $r_0 = 12 \pi a_{\rm dd}$. }
	\label{fig:phase_diagram_var}
\end{figure}

We start by evaluating the effect of temperature on the phase diagram in the thermodynamic limit, for which the trapping potential reads $U({\bf r}) = \frac{1}{2} m \omega_z^2 z^2$.
The trapping strength is fixed to $\hbar \omega_z/\epsilon_{\rm dd} =0.08 $.
We show in Fig.~\ref{fig:phase_diagram_var}, the phase diagram for a dipolar condensate with pancake geometry for $k_{\rm B} T=0$ (blue dashed lines) and $k_{\rm B} T/\epsilon_{\rm dd}=2$ [red solid lines (variational result) and markers (TeGPE solution)]. The latter corresponds to $T = 87$nK for a system of $^{164}$Dy atoms. The 2D condensate density is defined as $\rho_{\rm 2D} = \int dz \abs{\psi({\bf r})}^2$, where $\psi({\bf r})$ corresponds to the solution of Eq.~(\ref{TeGPE}) normalised to the particle number $N = \int d{\bf r} \abs{\psi({\bf r})}^2$.

Comparing the variational and numerical results for the finite temperature case in Fig.~\ref{fig:phase_diagram_var}, we note that the variational analysis captures the qualitative physics reasonably well, and thus represents a comparably inexpensive tool for its exploration.
The full numerical solution of Eq.~(\ref{TeGPE}) essentially amounts to a shift in both scattering length and density.

At zero temperature (blue lines), the dashed lines indicate a first-order phase transition between the fluid-solid, fluid-honeycomb and honeycomb-stripe phases. They converge to a point at which the phase transition is of second order~\cite{zhang19}. This qualitative phenomenology and qualitative shape of the phase diagram remains true at finite temperatures (red lines) as well.

We will focus the discussion now on the second-order point.
We note that a slight shift of the second-order point towards larger values of the scattering length occurs in the finite temperature case. This shift corresponds to $\Delta a_{\rm s}\simeq 3.3 a_0$ for $^{164}$Dy atoms, with $a_0$ denoting the Bohr radius. 
The most striking point when comparing the zero and finite temperature phase-diagram is the significant shift in condensate density. For the finite-temperature case the density of the second-order point is more than halved, $\rho_{\rm 2D}^{\rm c}(87$nK$)/\rho_{\rm 2D}^{\rm c}(0) = 0.46$ (note the different scales for the zero and finite-temperature case in Fig.~\ref{fig:phase_diagram_var}).
This reduction in density in the thermodynamic limit is promising for the realization of these phases in an experiment, since according to the estimate in Ref.~\cite{smith23} the lifetime due to three-body losses scales like $t_3 \sim 1/n^2$. Thus, in our case the lifetime could be expected to increase roughly by a factor of $\simeq 4.7$. We explore the effect of finite temperatures in the dynamical formation of these phases further in Sec.~\ref{sec:dynamics}.

\begin{figure}[b]
	\centering
	\includegraphics[width=\columnwidth]{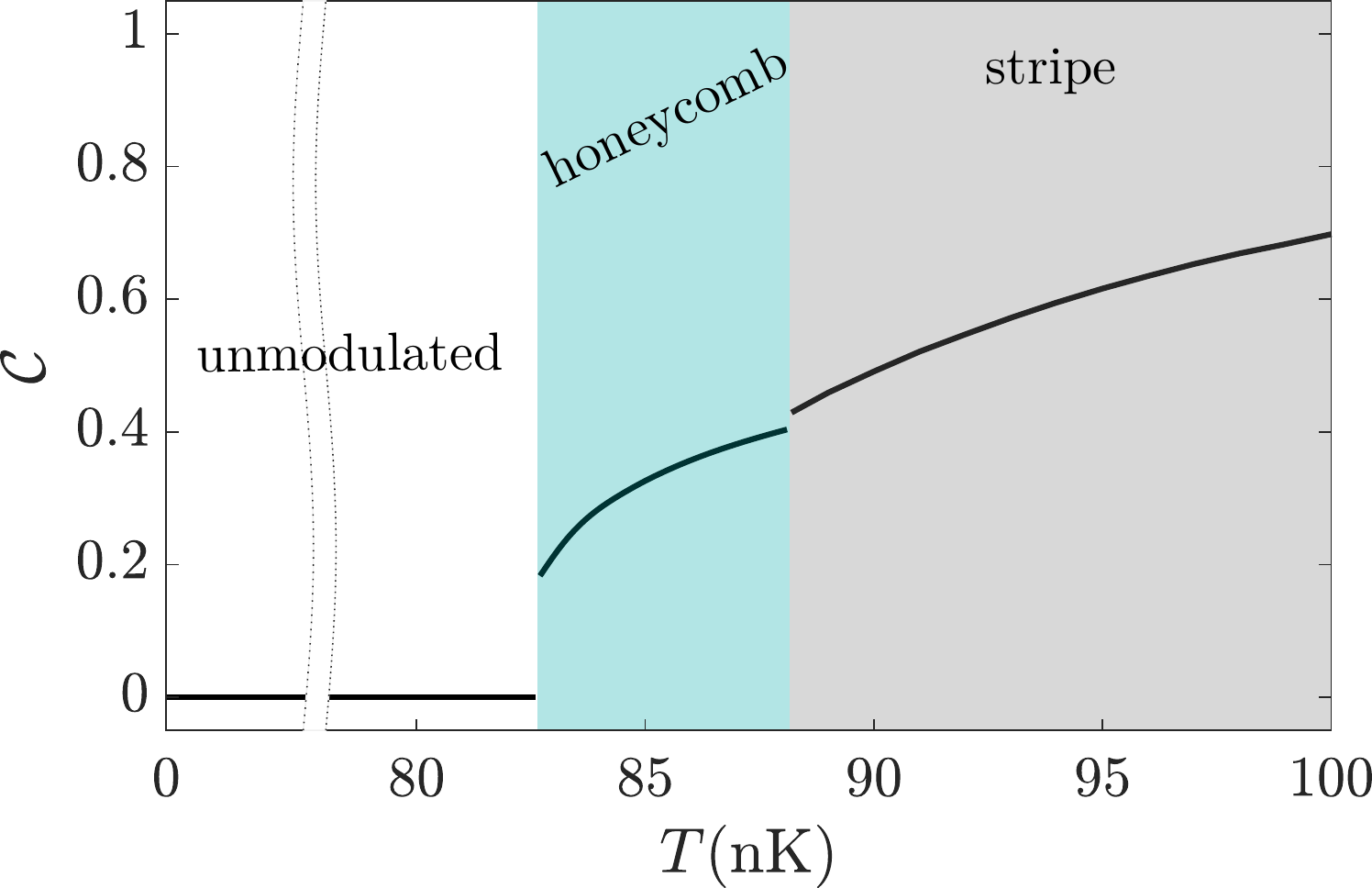}
	\caption{Temperature-driven supersolidity, visualised by showing the contrast of the ground state wave function (see Eq.~\ref{contrast}) as a function of temperature for $\rho_{\rm 2D} r_0^2=105$ and $a_{\rm s}/a_{\rm dd}=0.807$. }
\label{fig:contrast}
\end{figure}

These observations are consistent with previous results~\cite{baena22,baena24}. Yet, in this case the shift in density is substantially larger, highlighting that  dimensionality plays an important role. The increase of temperature for a given condensed density can promote a phase transition from the fluid phase to a modulated state. Fig.~\ref{fig:contrast} provides an example of the ground state phase transitions driven by temperature for $\rho_{\rm 2D} r_0^2=105$ and $a_{\rm s}/a_{\rm dd}=0.807$. It depicts the contrast of the ground state
\begin{equation}
 \mathcal{C}=\frac{|\psi(z=0)|^2_{\rm max}-|\psi(z=0)|^2_{\rm min}}{|\psi(z=0)|^2_{\rm max}+|\psi(z=0)|^2_{\rm min}}
 \label{contrast}
\end{equation}
at a fixed density. We see that when the temperature surpasses $\sim 83$nK, the honeycomb emerges as ground state with a finite contrast undergoing a first-order phase-transition. If we further increase the temperature beyond $\sim 88$nK, the honeycomb becomes energetically less favorable as compared to the stripe phase.

\begin{figure}[t]
    \centering
    \includegraphics[width=\columnwidth]{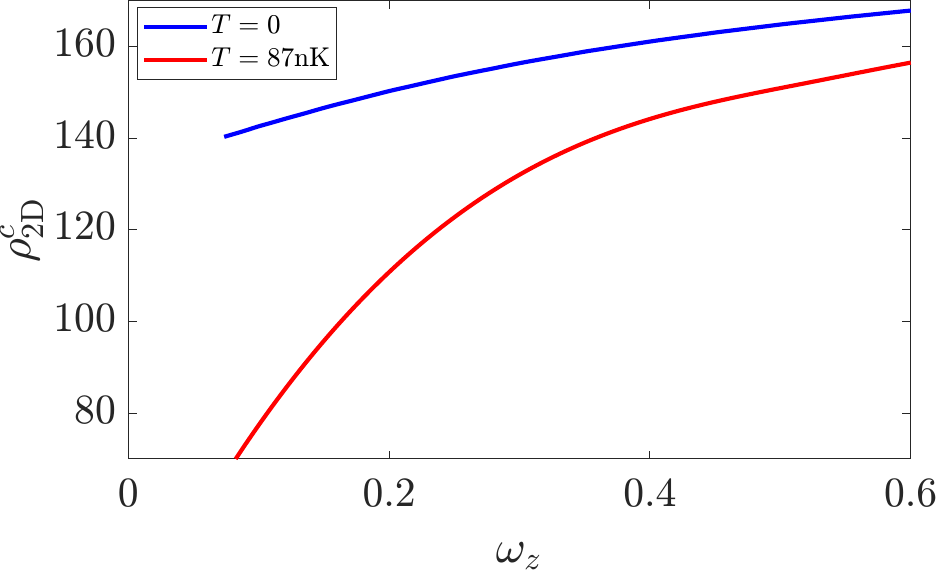}
    \caption{Dependence of the critical 2D density as function of the trapping frequency for $T=0$ (blue solid line) and $k_{\rm B} T/\epsilon_{\rm dd}=2$  (red solid line). The density and the harmonic frequency are expressed in units of $1/r_0^2$ and $\epsilon_{\rm dd}/\hbar$, respectively. }
    \label{fig:critical_omega}
\end{figure}

Let us now explore how varying the trapping frequency $\omega_z$
quantitatively changes the shift in density of the critical point.
Previous works at zero temperature show there is a competition between the peak densities and the particle number in the fully trapped system~\cite{zhang21}. Employing parameters that yield the honeycomb, labyrinth and stripe structures at densities for which the three-body losses are moderate for experiments involve prohibitively large condensed particle numbers ($N \sim 10^6$) and vice-versa, employing lower particle numbers ($N \sim 10^5$) leads to peak densities larger than $10^{15}{\rm cm}^{-3}$.

Fig.~\ref{fig:critical_omega} depicts the the critical density $\rho_{\rm 2D}^{\rm c}$ as a function of $\omega_z$ for zero and finite temperature with $k_{\rm B} T/\epsilon_{\rm dd} =  2$. The figure shows that the difference in the critical density of the second-order point for a finite temperature decreases as the frequency increases. This behaviour stems from the density dependence of the quantum and thermal fluctuation terms of Eqs.~(\ref{QF},~\ref{TF}). It has already been established that thermal fluctuations decrease upon increasing density while quantum fluctuations follow the opposite trend~\cite{baena22,baena24}.

\begin{figure}[t]
    \centering
    \includegraphics[width=\columnwidth]{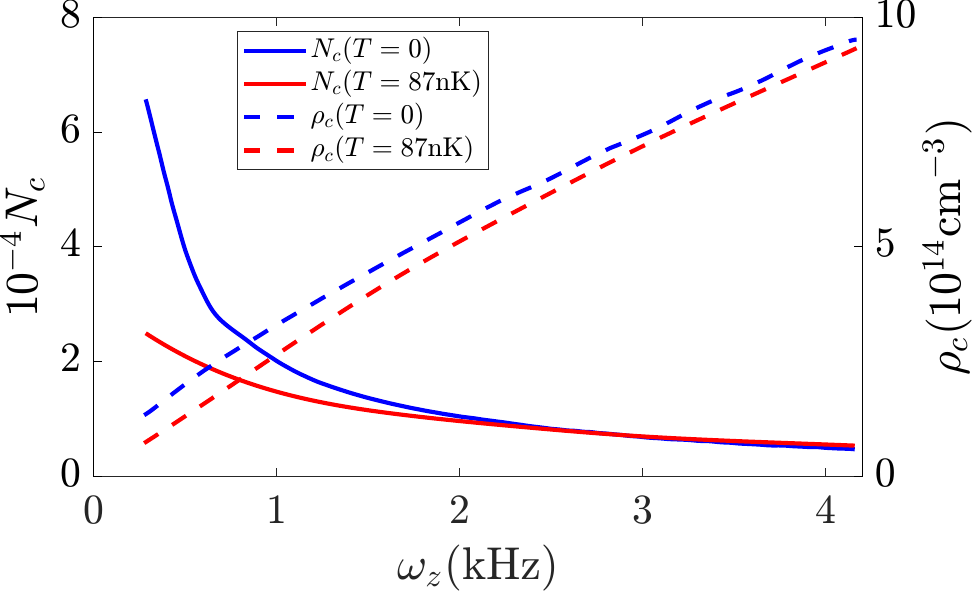}
    \caption{3D critical density (solid lines) and number of condensed particles per unit cell at the critical point (dashed lines) as a function of the trapping frequency for $T=0$ (blue) and $T = 87$ nK (red) for a gas of $^{164}$Dy atoms. }
    \label{fig:rho_omega}
\end{figure}

To further elucidate this point, Fig.~\ref{fig:rho_omega} shows the dependence of both the 3D critical density $\rho_{\rm c}=3\rho^c_{\rm 2D}/(4\sigma_z)$ and the number of condensed particles per unit cell at the critical point, $N_c=2\rho^c_{\rm 2D}\lambda^2/\sqrt{3}$, as a function of the trapping strength $\omega_z$ for both $k_{\rm B} T/\epsilon_{\rm dd} = 0 \text{ and } 2$. Here, the length $\lambda=2\pi/p$ is given by the wave vector of the modulated density at the critical point. Using $N_c$ we can estimate how many particles are required for a given number of unit cells of a density modulated state in the fully trapped system. Again, we note that temperature reduces $N_c$ significantly for small trapping frequencies.

\begin{figure}[t]
    \centering
    \includegraphics[width=0.5\textwidth]{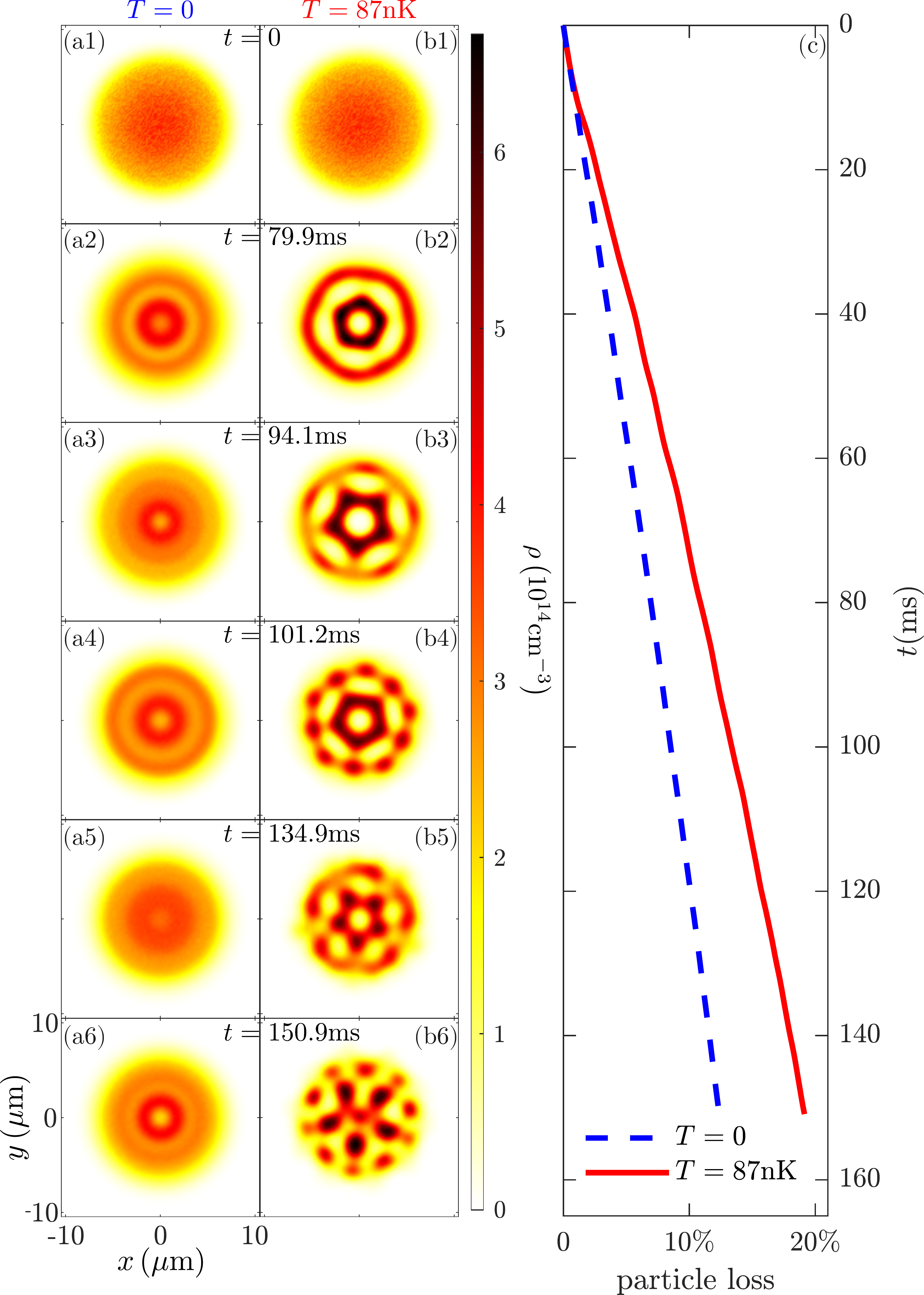}
    \caption{Subsequent dynamics after an interaction quench from $a_{\rm s}/a_{\rm dd} = 0.8$ to $0.757$ at $T=0$nK (a) and $T=87$nK (b). The trapping frequencies are $\hbar \omega_z / \epsilon_{\rm dd} = 0.11$ and $\hbar \omega_{\perp} / \epsilon_{\rm dd} = 0.052$. Initially, the number of condensed particles corresponds to $N=350000$. Panel (c) shows the decrease of the atom number due to the three-body losses with a loss-rate of $L_3 = 1.5 \times 10^{-41}$ m$^6$/s~\cite{bottcher19}.
    }
    \label{fig:honeycomb_dynamics_1}
\end{figure}

\begin{figure}[t]
    \centering
    \includegraphics[width=0.5\textwidth]{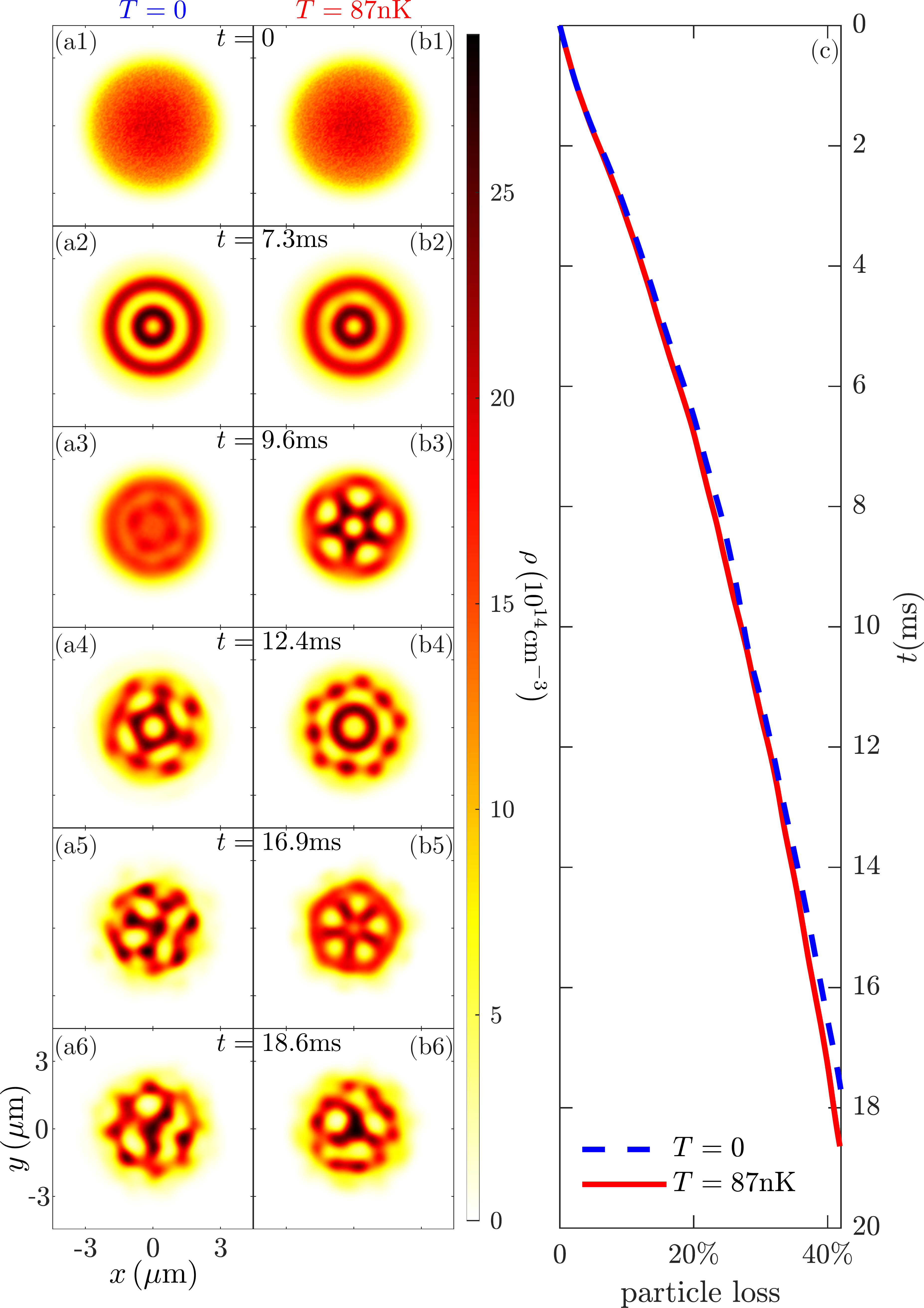}
    \caption{Dynamics following an interaction quench from $a_{\rm s}/a_{\rm dd} = 0.65$ to $0.595$ at $T=0$nK (a) and $T=87$nK (b). The trapping frequencies are $\hbar \omega_z / \epsilon_{\rm dd} = 0.75$ and $\hbar \omega_{\perp} / \epsilon_{\rm dd} = 0.32$. The initial number of condensed particles is set to $N=100000$. Panel (c) shows the decrease of the atom number due to the three-body losses with a loss-rate of $L_3 = 1.5 \times 10^{-41}$ m$^6$/s~\cite{bottcher19}.
    }
    \label{fig:honeycomb_dynamics_2}
\end{figure}

In view of the recent progress in the control and reduction of reactive losses for ultracold dipolar  molecules~\cite{Gorshkov08,Junyu2023,Karam2023,Bigagli2023,Mukherjee2024,Bigagli2024} and the realization of the first molecular dBEC~\cite{Bigagli2024}, it is interesting to put the previous results in the context of molecules. Due to their considerably larger dipole moment as compared to dBECs of a single species, the reduced temperature of $k_{\rm B} T/\epsilon_{\rm dd} = 2$ corresponds to $T = 1$nK for a gas of NaCs molecules.
For this extremely low temperature the values of $N_c$ and $\rho_{\rm c}$ lay in the range $N_c \in (0.5, 2.5) \times 10^4$ and $\rho_{\rm c} \in (10^{11}, 10^{12})$cm$^{-3}$ for trapping frequencies $\omega_z \in (3, 40)$ Hz. While the values of the critical density are less or equal compared to those in the recent experiment of Ref.~\cite{Bigagli2024}, the number of condensed particles per unit cell greatly exceeds that of the experimental condensate, which consists of a few hundred atoms.
That being said, the production of molecular dBECs is still at an early stage and future developments might lead to molecular dBECs with higher particle numbers.

\section{Real-time excitation of a honeycomb state in a dipolar BEC}
\label{sec:dynamics}

Thus far we restricted our discussion to phases in the thermodynamic limit. We focus our attention now on the fully trapped system. For that matter, we run real-time simulations of the TeGPE to model the experimental realization of the honeycomb state for a system of $^{164}$Dy atoms following a quench of the scattering length at $T = 0$ and $T = 87$nK. We have included three-body losses in the same way as in Refs.~\cite{bottcher19,zhang21}. The results are shown in Figs.~\ref{fig:honeycomb_dynamics_1} and \ref{fig:honeycomb_dynamics_2}.

For a trapping strength of $\hbar \omega_z / \epsilon_{\rm dd} = 0.11$ and $N=350000$ condensed atoms, a honeycomb state of a large lifetime of $\sim 40$ms with a moderate peak density of $\rho_{\rm peak} \sim 6 \times 10^{14}$cm$^{-3}$ is displayed in Fig.~\ref{fig:honeycomb_dynamics_1}. We can see from the results that temperature favours the formation of the honeycomb state, while the calculation at zero temperature does not lead to a modulated state. Unfortunately, decreasing the particle number under these conditions further results in the disappearance of the honeycomb structure. To decrease the condensed particle number while retaining the honeycomb state we have to set a tighter confinement along the $z$ axis, which will increase the density. We observe a structure with a much shorter lifetime of $\sim 7$ms for $\hbar \omega_z / \epsilon_{\rm dd} = 0.75$ and $N=100000$ condensed atoms (Fig.~\ref{fig:honeycomb_dynamics_2}), however at the cost of a considerably larger peak density $\rho_{\rm peak} \sim 3 \times 10^{15}$cm$^{-3}$. Thermal effects here have a smaller impact than in the prior case, as thermal fluctuations dominate at smaller densities as already discussed~\cite{baena22, baena24}.

Summarisingly, the two cases we presented have the purpose to portray two extreme parameter domains, one case with a large particle number and a comparably small density that is strongly affected by temperature and one case with a comparably small particle number and larger density that is less affected.
In the first case the particle number is significantly reduced as compared to the  zero-temperature situation~\cite{zhang21}.

\section{Conclusions}

In this paper we have explored whether thermal fluctuations might assist in promoting pattern-formation to such an extent that the high-density physics of a dBEC with pancake symmetry becomes experimentally accessible. This includes access to novel phases such as honeycomb and stripe phases as well as the second-order point of the phase diagram. We have found that an increase in temperature indeed can lead to a significant decrease in the necessary density to probe the high-density physics of the flattened dBEC. We have also shown real-time simulations with realistic interaction quenches that gave rise to the formation of a honeycomb.
Thus, we conclude that temperature indeed might present a promising route towards the potential realization of these high-density phases.

Beyond probing the high-density physics of dBECs and pattern-formation, this work might pave a further pathway towards exploring finite temperature effects in dBECs due to the clear signature of the emerging patterns.
Furthermore, higher-order theories beyond what has been presented here could, for instance, quantitatively study the effect of temperature on the superfluid properties of the density modulated structures, like the honeycomb or the stripe~\cite{hertkorn21,gallemi22,zhang24}.
For this purpose, \textit{ab-initio} methods, like Monte Carlo algorithms represent an excellent option. Such methods would be able to study the regime of even higher temperatures than what was considered here, where the system is mostly non-condensed.

\section{Acknowledgements}

This work was supported by National Key Research and Development Program of China (Grant No.: 2021YFA1401700), the National Nature Science Foundation of China (Grant No.: 12104359), Shaanxi Academy of Fundamental Sciences (Mathematics, Physics) (Grant No.: 22JSY036). J.S-B acknowledges support by the Spanish Ministerio de Ciencia e Innovaci\'on (MCIN/AEI/10.13039/501100011033, grants PID2020-113565GB-C21 and PID2023-147469NB-C21), and by the Generalitat de Catalunya (grant 2021 SGR 01411). Y.C.Z. acknowledges the support of Xi'an Jiaotong University through the ``Young Top Talents Support Plan" and Basic Research Funding as well as the High-performance Computing Platform of Xi'an Jiaotong University for the computing facilities.

\bibliography{references}

\bibliographystyle{apsrev4-1}

\end{document}